\documentclass[12pt,reqno]{iopconfser}

\usepackage{amsfonts,amsmath,amssymb,inputenc}

\begin{document}

\def\Nnum{\mathbb{N}}
\def\Rnum{\mathbb{R}}
\def\Cnum{\mathbb{C}}
\def\Re{\mathrm{Re}\,}
\def\Im{\mathrm{Im}\,}

\def\Hnum{\mathbb{H}}
\def\Qnum{\mathbb{Q}}
\def\Onum{\mathbb{O}}

\def\u{\mathbf{u}}
\def\F{\mathbf{F}}
\def\L{\mathbf{L}}
\def\M{\mathbf{M}}
\def\Psi{\boldsymbol{\psi}}

\def\e{\mathbf{e}}
\def\octa{\mathbf{a}}
\def\octb{\mathbf{b}}
\def\octc{\mathbf{c}}

\def\mul#1{{\mathrm #1}}

\def\Ref#1{Ref.~\cite{#1}}

\newtheorem{prop}{Proposition}[section]

\numberwithin{equation}{section}
\tolerance=50000

\allowdisplaybreaks[3]

\title{A search for integrable evolution equations\\ with Lax pairs over the octonions}

\author{Stephen C. Anco$^{1}$, Philic Lam$^{1}$, Thomas Wolf$^{1}$}

\affil{$^{1}$Department of Mathematics \& Statistics, Brock University, St Catharines, Canada}

\email{sanco@brocku.ca, hinho1106@gmail.com, twolf@brocku.ca}

\begin{abstract}
Four new integrable evolutions equations with operator Lax pairs are found 
for an octonion variable. 
The method uses a scaling ansatz to set up a general polynomial form 
for the evolution equation and the Lax pair, using KdV and mKdV scaling weights. 
A condition for linear differential operators to be a Lax pair over octonions 
is formulated and solved for the unknown coefficients in the polynomials. 
\end{abstract}

\section{Introduction}\label{sec:intro}

Very well-known prototypical examples of integrable evolution equations \cite{AblCla-book}
are the Korteweg-de Vries (KdV) equation $u_t = uu_x + u_{xxx}$ 
and the modified Korteweg-de Vries (mKdV) equation $u_t = \sigma u^2u_x + u_{xxx}$
for a real scalar field $u$, up to a scaling transformation on $(t,x,u)$
where $\sigma=\pm1$. 
These two equations are linked by a Miura transformation
and possess the full slate of integrability properties:
higher symmetries and conservation laws; 
bi-Hamiltonian formulation; 
Hirota bilinear form; 
Darboux/auto-Backlund transformation; 
Lax pair; 
inverse scattering transform (IST);
multi-soliton solutions. 

There has been longstanding interest in finding and studying 
integrable multi-component generalizations of these equations
\cite{Sok-book}. 
Known types comprise two scalar fields $(u_1,u_2)$, 
a vector field $\vec u$ in the mKdV case 
and a scalar field $u$ coupled to a vector field $\vec u$ in the KdV case, 
as well as a matrix field $\mathrm u$ for both KdV and mKdV cases, 
which includes a quaternion field as a special case. 
In all of these examples, 
their integrability is connected to Jordan triple systems \cite{SviSok,SheSok}
which are based on commutative, non-associative algebras that obey 
a certain partial-associativity law. 

A natural question is to seek a further generalization for an octonion field $\u$. 
Octonions constitute an 8-dimensional normed division algebra that is non-commutative as well as non-associative 
(see \Ref{Conway-book,Baez} for a full explanation of its definition properties). 
No connection between the octonion algebra and Jordan algebras/triple systems is known. 

Some recent work \cite{ResSotVei,FerResSot} has found an octonion KdV equation
$\u_t = \tfrac{1}{2}(\u^2)_x + \u_{xxx}$
and presented a Lax pair for it. 
This example motivates undertaking a systematic search for integrable octonion evolution equations with Lax pairs. 

The results in the present paper consist of 
a second Lax pair for the octonion KdV equation,
plus four types of new integrable octonion equations:\\
$\bullet$
a potential-KdV equation with two Lax pairs; \\
$\bullet$
a one-parameter family of mKdV equations with three Lax pairs;\\
$\bullet$
a mKdV-like equation whose nonlinear term is $[[\u,\u_x],\u]$,
with three Lax pairs;\\
$\bullet$
a novel variant of the previous two types that has an additional $[\u,\u_{xx}]$ term,
with three Lax pairs. 

The method that we use is a generalization of a computational 
scaling ansatz method \cite{HicHerLarGok} using undetermined coefficients 
to find Lax pairs for given scalar integrable equations that are scaling homogeneous. 
There are two main aspects to our version: 
a scaling ansatz is used to search for both the Lax pair and the octonion equation together; 
the Lax pair is re-formulated so as to allow the field variable to lack commutativity and associativity. 

A computational difficulty is that the octonions (and quaternions) possess certain product identities 
which lead to free parameters multiplying terms that collectively are equal to zero 
in the multi-component system and its Lax pair. 
We remove these terms by finding and utilizing the relevant identities a priori. 

In Section~\ref{sec:method}, 
we outline the steps in the method. 
We present the integrable octonion evolution equations and their Lax pairs 
in Sections~\ref{sec:KdVscaling} and~\ref{sec:mKdVscaling}. 
Some concluding remarks are made in Section~\ref{sec:conclude}. 

Throughout, bold-face symbols denote octonion quantities; 
$\e_0$ will denote the real basis element in the octonion algebra,
and the imaginary elements will be denoted $\e_1,\ldots,\e_7$. 
Octonion product identities need for the results are shown in the Appendix.

\section{Outline of Method}\label{sec:method}

Recall that a Lax pair for a scalar evolution equation $u_t=F[u]$
consists of a pair of linear differential operators $L$ and $M$ in $\partial_x$, 
with coefficients depending on $u$ and $x$-derivatives of $u$, 
such that 
\begin{equation}\label{laxpair}
L_t = [M,L]
\end{equation} 
holds for all solutions of the evolution equation. 
A Lax pair \eqref{laxpair} is the compatibility condition for the linear system 
\begin{subequations}\label{lax.eqns}
\begin{align}
L\psi & = \lambda \psi,
\label{lax.eqn.L}
\\
\psi_t & = M\psi , 
\label{lax.eqn.M}
\end{align}
\end{subequations}
where $\lambda$ is a constant called the spectral parameter. 
This is the starting point for the IST, 
which takes equations \eqref{lax.eqns} to be a linear scattering problem
with eigenfunctions $\psi$ and corresponding eigenvalues $\lambda$. 
Since the eigenvalues are time-independent, the linear system is called isospectral. 

To preserve this structure for integrable octonion equations,
we need to handle the non-commutativity and non-associativity of expressions 
involving $u$, $\psi$, and their derivatives. 
The most straightforward way is to reformulate the Lax pair 
at the level of the corresponding isospectral linear system. 
Observe that the operator equation \eqref{laxpair} is equivalent to 
$L_t\psi = M(L\psi) - L(M\psi)$, 
and thus 
$M(L\psi) = L_t\psi + L(M\psi) = (L\psi)_t$ holds by equation \eqref{lax.eqn.M}. 
Moreover, use of equation \eqref{lax.eqn.L} gives 
$M(L\psi)=M(\lambda \psi)= \lambda M\psi$,
from which we have $(L\psi)_t = \lambda M\psi$. 
Substituting equation \eqref{lax.eqn.L} into the left-hand side directly gives 
$(L\psi)_t  = \lambda_t\psi + \lambda M\psi$,
and hence subtracting the right-hand side yields $\lambda_t=0$. 
Therefore, we have the following result. 

\begin{prop}\label{prop:laxpair}
A Lax pair \eqref{laxpair} is equivalent to the linear system 
\begin{equation}\label{laxpair.sys}
(L\psi)_t = M(L\psi),
\quad
\psi_t = M\psi . 
\end{equation}
This system augmented by $L\psi=\lambda\psi$ is isospectral, namely
$\lambda_t =0$. 
\end{prop}

This formulation of a Lax pair is very useful because it does not assume 
any commutativity or associativity properties for $u$, $\psi$, and their derivatives. 
In particular, it holds assuming only that $\lambda$ distributes through products of those functions. 
As a consequence, we can adopt the system \eqref{laxpair.sys} 
as the defining condition for Lax integrability of octonion evolution equations:
\begin{gather}
\u_t =\F[\u],
\label{evol.eqn}
\\
(\L\Psi)_t|_{\u_t=\F[\u]} = \M(\L\Psi),
\quad
\Psi_t = \M\Psi
\label{laxpair.cond}
\end{gather}
where $\u(x,t)$ and $\Psi(x,t)$ are octonion functions. 

Now, suppose that we have a polynomial ansatz for $\F[\u]$, $\L\Psi$ and $\M\Psi$,
with unknown real constant coefficients. 
Then the condition \eqref{laxpair.cond} provides a determining system 
on these coefficients as follows. 
Firstly, we expand 
\begin{equation}\label{u.psi.basis}
\u = u_0 \e_0 + u_1 \e_1 + \ldots + u_7 \e_7
\quad\text{ and }\quad
\Psi = \psi_0 \e_0 + \psi_1 \e_1 + \ldots + \psi_7 \e_7
\end{equation}
(and their derivatives) in terms of octonion basis elements $\{\e_0,\e_1,\ldots,\e_7\}$,
where 
\begin{equation}\label{comps}
u_0,u_1,\ldots,u_7
\quad\text{ and }\quad
\psi_0,\psi_1,\ldots, \psi_7
\end{equation}
are real scalar functions. 
Secondly, we split the two equations \eqref{laxpair.cond} 
with respect to the octonion basis elements, giving two systems of 8 equations,
each of which we further split in the jet space of the scalar functions \eqref{comps}.
These splittings yield an overdetermined system of algebraic equations 
on the unknown constant coefficients in the ansatze. 
Finally, we solve this algebraic system, taking into account all case splittings that arise,
with $\L\Psi$ and $\M\Psi$ required to have some dependence on $\u$,
and with $\F[\u]$ required to have nonlinear dependence on $\u$ (so as to exclude linear evolution equations). 

A polynomial ansatz will be generated by imposing scaling homogeneity. 
We will carry this out first for KdV scaling weights, 
and next for mKdV scaling weights. 

There two important well-known remarks about gauge freedom in a Lax pair $(L,M)$.
First, $(L^p,M)$ is a Lax pair for any positive integer $p$. 
The proof is a straightforward use of induction. 
Second, $(L,M+aL^p)$ is a Lax pair for any constant $a$ and any positive integer $p$. 
This is a simple consequence of $[L,L^p] =0$.

\section{KdV scaling-weight equations}\label{sec:KdVscaling} 

The scalar KdV equation $u_t = uu_x + u_{xxx} =F[u]$ possesses 
the group of scaling symmetries $t\to \mu^{-3} t$, $x\to \mu^{-1} x$, $u\to \mu^{2}u$, 
with group parameter $\mu \neq 0$, 
where the scaling weight is $w_F= 5$. 
A standard Lax pair is given by the linear differential operators 
$L=\partial_x^2 + \tfrac{1}{6}u$
and 
$M= 4\partial_x^3 + u\partial_x +\tfrac{1}{2}u_x$, 
which satisfy the Lax equation \eqref{laxpair} 
when $u$ is any solution of the KdV equation. 
Their scaling weights are $w_L= 2$ and $w_M = 3$. 

The search for octonion generalizations comprises three main steps. 

Step (1): 
We first set up an ansatz for an octonion evolution equation with the scaling weights
$w_x=-1$, $w_t=-3$, $w_\u = 2$, $w_\F = 5$:
\begin{equation}\label{kdv.F}
\u_t = c_1 \u \u_x + c_2 \u_x \u + c_3 \u_{xxx} =\F[\u]
\end{equation}
which is the most general polynomial having the same scaling weights as the real KdV equation,
where $c_1$, $c_2$, $c_3$ are real constants. 
Note that we can put 
\begin{equation}
c_3=1
\end{equation}
via a scaling transformation on $t,x$. 

Next we write down the ansatz for the Lax pair, 
using an octonion function $\Psi$ defined to have scaling weight $0$.
For scaling weights $w_\L = 2$ and $w_\M = 3$, 
the most general polynomials (linear in $\Psi$) are respectively given by 
\begin{equation}\label{kdv.Lpsi}
\L\Psi=\Psi_{xx} + l_1 \u\Psi + l_2 \Psi \u
\end{equation}
where $l_1$, $l_2$ are real constants, 
and 
\begin{equation}\label{kdv.Mpsi}
\M\psi= m_1 \Psi_{xxx} + m_2 \u\Psi_x + m_3 \Psi_x \u + m_4 \u_x\Psi + m_5 \Psi \u_x
\end{equation}
where $m_1$, $\ldots$, $m_5$ are real constants. 

Step (2): 
We first substitute expressions \eqref{kdv.F}, \eqref{kdv.Lpsi} and \eqref{kdv.Mpsi}
into the Lax pair equations \eqref{laxpair.cond},
and next substitute the basis expansions \eqref{u.psi.basis} for $\u$ and $\Psi$.  
Then we split these two equations with respect to the octonion basis,
and split again with respect to the jet variables of the components \eqref{comps}. 
This has been carried out in Maple, using the octonion algebra 
available in 'AlgebraLibraryData' (as part of the 'LieAlgebras' package). 
The splitting yields an overdetermined system of 27 algebraic equations
for the 9 real constants $c_1$, $c_2$, $l_1$, $l_2$, $m_1$, $\ldots$, $m_5$. 
Note that the system is quadratically nonlinear. 

Step (3):
We solve the system in Maple by 'rifsimp' (in the 'DEtools' package), 
with the following priority chosen for the unknowns: 
$\{m_1,\ldots,m_5\}$ (highest); $\{l_1,l_2\}$ (next highest); $\{c_1,c_2\}$ (lowest). 
This priority ensures that the final case tree of solutions is parameterized in terms of 
the constant coefficients in the evolution equation \eqref{kdv.F}. 
The case tree comprises any necessary case splittings needed in solving for the unknowns. 

Two solutions are obtained: 
\begin{subequations}
\begin{align}
& 
c_1=c_2=\alpha,\ 
l_1=\tfrac{1}{3} \alpha,\  
m_1=4,\ 
m_2=2\alpha,\ 
m_4=\alpha,\ 
l_2=m_3 = m_5=0;
\\
& 
c_1=c_2=\alpha,\ 
l_2=\tfrac{1}{3} \alpha,\  
m_1=4,\ 
m_3=2\alpha,\ 
m_5=\alpha,\ 
l_1=m_2 = m_4=0 
\end{align}
\end{subequations}
where $\alpha\neq0$ is a free parameter.

\subsection{Results}

Both solutions yield the same octonion evolution equation 
\begin{equation}\label{kdv.oct.eqn}
\u_t = \alpha(\u \u_x + \u_x \u) + \u_{xxx} , 
\end{equation}
and two Lax pairs
\begin{equation}\label{kdv.oct.laxpair1}
\L\Psi=\Psi_{xx} + \tfrac{1}{3}\alpha \u\Psi , 
\quad
\M\Psi= 4 \Psi_{xxx} + 2\alpha \u\Psi_x +\alpha\u_x\Psi , 
\end{equation}
and
\begin{equation}\label{kdv.oct.laxpair2}
\L\Psi=\Psi_{xx} + \tfrac{1}{3}\alpha \Psi\u , 
\quad
\M\Psi= 4 \Psi_{xxx} +2\alpha\Psi_x\u +  \alpha\Psi \u_x . 
\end{equation}
Factoring out $\Psi$ from $\L\Psi$ and $\M\Psi$ in each Lax pair yields operators 
$\L = \partial_x^2 + \tfrac{1}{3}\alpha \u$
and $\M= 4 \partial_x^3  +2\alpha\u\partial_x +  +\alpha\u_x$
which act on $\Psi$ by either left or right multiplication. 

The real reduction, in which $\u$ is replaced by $u$ and $\Psi$ is replaced by $\psi$,
is given by the real KdV equation and its Lax pair with weight $w_L = 2$, 
after a scaling transformation of $\u$ to put $\alpha=\tfrac{1}{2}$.

It is interesting to repeat the search using higher weights $w_\L$,
since this could possibly lead to other evolution equations with KdV scaling weights. 
For $w_\L=3$, no solution exists. 
Two solutions are found for $w_\L =4$. 
They give the octonion KdV equation \eqref{kdv.oct.eqn}
with the previous operator $\M$, 
and $\L_{(2)} = \L^2$ given by the square of the previous operator $\L$.

\section{mKdV scaling-weight equations}\label{sec:mKdVscaling}

The scalar mKdV equation $u_t = \sigma u^2u_x + u_{xxx} =F[u]$, $\sigma^2=1$, 
possesses the group of scaling symmetries 
$t\to \mu^{-3} t$, $x\to \mu^{-1} x$, $u\to \mu u$, 
with group parameter $\mu \neq 0$, 
where the scaling weight is $w_F= 4$. 
It's lowest weight Lax pair is 
$L=\partial_x +k u$
and 
$M= k( u_{xx} -\tfrac{1}{3}\sigma u^3 )$, 
which satisfies the Lax equation \eqref{laxpair} 
when $u$ is any solution of the mKdV equation,
where $k\neq0$ is an arbitrary constant. 
The scaling weights of these linear differential operators are $w_L= 1$ and $w_M = 3$. 

The search for octonion generalizations goes as follows. 

Step (1): 
We set up an ansatz for an octonion evolution equation with the scaling weights
$w_x=-1$, $w_t=-3$, $w_\u = 1$, $w_\F = 4$:
\begin{equation}\label{mkdv.F}
\u_t = c_1 \u^4 + c_2 \u^2 \u_x + c_3 \u_x \u^2 + c_4 \u \u_x \u 
+ c_5 \u_x^2 +c_6 \u_{xx}\u + c_7 \u \u_{xx} +c_ 8
 \u_{xxx} 
=\F[\u]
\end{equation}
which uses the associativity property \eqref{prod.id1} 
of products involving at most two octonion variables, 
whereby we have 
$(\u^2)\u_x = \u(\u\u_x)$, 
$\u_x (\u^2) = (\u_x\u)\u$, 
and $(\u \u_x) \u = \u(\u_x\u)$,
plus $\u(\u\u^2) = \u (\u^2 \u) = \u^2 \u^2 = (\u \u^2) \u = (\u^2 \u) \u$. 
Thus, $\F[\u]$ is the most general polynomial having the same scaling weights as the real mKdV equation,
where $c_1$, $\ldots$, $c_7$ are real constants. 
Note that we can put 
\begin{equation}
c_8 =1
\end{equation}
via a scaling transformation on $t,x$. 

The ansatz for a lowest-weight Lax pair is given by 
\begin{equation}\label{mkdv.Lpsi}
\L\Psi=\Psi_x + l_1 \u\Psi + l_2 \Psi \u
\end{equation}
where $l_1$, $l_2$ are real constants, 
and 
\begin{equation}\label{mkdv.Mpsi}
\begin{aligned}
\M\psi &= m_1 \Psi_{xxx} 
+m_2 \u \Psi_{xx} + m_3 \Psi_{xx} \u
+m_4 \u_x \Psi_x + m_5 \Psi_x \u_x 
\\&\qquad
+ m_6 \u^2 \Psi_x + m_7 \u \Psi_x \u +  m_8 \Psi_x \u^2 
+ m_{9} \u_{xx} \Psi + m_{10} \Psi \u_{xx}
\\&\qquad
+ m_{11} (\u\u_x) \Psi + m_{12} \u(\u_x \Psi) 
+ m_{13} (\u_x\u) \Psi + m_{14} \u_x(\u \Psi) 
\\&\qquad
+ m_{15} (\u \Psi)\u_x + m_{16} \u(\Psi \u_x) + m_{17} (\u_x \Psi)\u + m_{18} \u_x(\Psi \u) 
\\&\qquad
+ m_{19} (\Psi \u)\u_x + m_{20} \Psi(\u \u_x) + m_{21} (\Psi\u_x)\u + m_{22} \Psi (\u_x\u) 
\\&\qquad
+m_{23} \u^3 \Psi +m_{24} \u^2 \Psi \u +m_{25} \u \Psi \u^2 +m_{26} \Psi \u^3 
\end{aligned}
\end{equation}
where $m_1$, $\ldots$, $m_{26}$ are real constants. 
Here we have again used the associativity property \eqref{prod.id1}. 
Consequently, these expressions \eqref{mkdv.Lpsi} and \eqref{mkdv.Mpsi}
are the most general polynomials (linear in $\Psi$) having the respective scaling weights
$w_\L = 1$ and $w_\M = 3$. 
By use of the gauge freedom $(\L,\M)\to (\L,\M+a\L^3)$, with $a=-m_1$, 
we can put 
\begin{equation}\label{mkdv.elim1}
m_1=0 . 
\end{equation}
Additionally, 5 of the 12 terms involving products of $\u$, $\u_x$, $\Psi$ in $\M\Psi$ 
can be eliminated by means of the product identities \eqref{prod.id3}:
$\u(\u_x\Psi) -(\u\u_x)\Psi 
= (\u_x\u)\Psi -\u_x(\u\Psi)
= (\u\Psi)\u_x -\u(\Psi\u_x) 
= \u_x(\Psi\u) - (\u_x\Psi)\u
= \Psi(\u\u_x) - (\Psi\u)\u_x
= \Psi(\u_x\u) - (\Psi\u_x)\u$. 
This allows putting
\begin{equation}\label{mkdv.elim2}
m_{14}=m_{16}=m_{18}=m_{19}=m_{21}=0 . 
\end{equation}

Step (2): 
We substitute expressions \eqref{mkdv.F} and \eqref{mkdv.Lpsi}
into the Lax pair equations \eqref{laxpair.cond},
followed by expression \eqref{mkdv.Mpsi} with the terms \eqref{mkdv.elim1}--\eqref{mkdv.elim2} removed, 
and next we substitute the basis expansions \eqref{u.psi.basis} for $\u$ and $\Psi$. 
Using Maple, 
we split the Lair pair equations with respect to the octonion basis,
and split again with respect to the jet variables of the components \eqref{comps}, 
yielding an overdetermined system of 100 algebraic (quadratically nonlinear) equations
for the 29 real constants $c_1$, $\ldots$, $c_7$, $l_1$, $l_2$, $m_2$, $\ldots$, $m_{13}$, $m_{15}$, $m_{17}$, $m_{20}$, $m_{22}$, $\ldots$, $m_{26}$. 

Step (3):
We solve the system in Maple by 'rifsimp', 
with the following priority chosen for the unknowns: 
$\{m_i\}$ (highest); $\{l_j\}$ (next highest); $\{c_k\}$ (lowest). 
Six solutions are obtained. They yield two different families of evolution equations, 
each having three different Lax pairs.

\subsection{Results I}

The first family of octonion evolution equations is given by 
\begin{equation}
c_2 = c_3 = -3\alpha^2 + \beta,\ 
c_4 =\beta,\
c_1 = c_5 = c_6 = c_7 = 0 , 
\end{equation}
which yields
\begin{equation}\label{mkdv.fam1}
\u_t = (\beta -3\alpha^2)(\u^2 \u_x + \u_x \u^2) +\beta \u \u_x \u + \u_{xxx}
\end{equation}
where $\alpha$ and $\beta$ are free parameters. 
Its three Lax pairs are most easily expressed utilizing left and right multiplication operators 
\begin{equation}\label{L.R.mult}
\mul{L}(\octa)\octb = \octa\octb,
\quad
\mul{R}(\octa)\octb = \octb\octa
\end{equation}
combined with adding a suitable choice of product identities \eqref{prod.id3} on $\u$, $\u_x$, $\Psi$, 
yielding
\begin{equation}
\L=\partial_x^2 -\alpha \mul{K}(\u),
\quad
\M = \alpha \mul{K}(\u_{xx}) +\alpha^2[\mul{K}(\u),\mul{K}(\u_x)] +(\beta -2\alpha^2)\alpha \mul{K}(\u^3),
\end{equation}
where $\mul{K} = \mul{L}, \mul{R}, \mul{L} +\mul{R}$. 

To classify the types of evolution equations in this two-parameter family \eqref{mkdv.fam1}, 
it is useful first to look at the real reduction 
in which $\u$ and $\Psi$ are replaced by $u$ and $\psi$, 
giving $u_t = 3(\beta -2\alpha^2) u^2 u_x + u_{xxx}$. 
Two cases now arise, given by $\beta -2\alpha^2$ is zero or non-zero. 

In the case $\beta \neq 2\alpha^2$, 
the reduction yields the real KdV equation and its Lax pair with weight $w_L = 1$, 
up to a scaling transformation on $u$. 
Thus, putting $3(\beta -2\alpha^2) = \sigma$ by scaling $\u$, 
we have
\begin{equation}
\beta = 2\alpha^2 + \tfrac{1}{3}\sigma,
\end{equation}
which gives 
\begin{equation}\label{mkdv.case1}
\u_t = (2\alpha^2 +\tfrac{1}{3}\sigma)  \u \u_x \u  -(\alpha^2 -\tfrac{1}{3}\sigma)(\u^2\u_x + \u_x \u^2)  + \u_{xxx},
\quad
\alpha\neq 0 . 
\end{equation}
This is a one-parameter family of octonion mKdV equations. 
Special cases include: \\
$\u_t = \u \u_x \u  + \u_{xxx}$ for $\alpha = \tfrac{1}{\sqrt{3}}$ and $\sigma=1$; 
and 
$\u_t = -\tfrac{1}{2} \{\u^2,\u_x\}  + \u_{xxx}$ for $\alpha = \tfrac{1}{\sqrt{6}}$ and 
$\sigma=-1$,
in terms of an anticommutator (symmetric) product. 

In the case $\beta =2\alpha^2$, 
the real reduction is the linear Airy equation $u_t = u_{xxx}$. 
Taking 
\begin{equation}
\beta = 2, 
\quad
\alpha =1, 
\end{equation}
after scaling $\u$,  
we obtain 
\begin{equation}\label{mkdv.case2}
\u_t = [[\u,\u_x] \u]  + \u_{xxx}
\end{equation}
in terms of a nest commutator (antisymmetric) product. 
This is a novel type of octonion mKdV-type equation, 
which relies on a lack of commutativity.

\subsection{Results II}

The second family of octonion evolution equations is given by 
\begin{equation}
c_2 = c_3 = -6 \alpha^2 + \beta,\
c_4=\beta,\
c_6 = -c_7 = -\alpha,\
c_1 = c_5 = 0 , 
\end{equation}
which yields
\begin{equation}\label{mkdv.fam2}
\u_t = (\beta -6\alpha^2)(\u^2 \u_x + \u_x \u^2) +\beta \u\u_x\u +\alpha (\u \u_{xx} - \u_{xx} \u) + \u_{xxx}
\end{equation}
where $\alpha$ and $\beta$ are free parameters. 
When we add a suitable choice of product identities \eqref{prod.id3} 
on $\u$, $\u_x$, $\Psi$ 
to $\M\Psi$ in the three Lax pairs, 
they can be expressed as 
\begin{gather}
\L=\partial_x^2 -\alpha \mul{K}(\u),
\\
\M = \alpha \mul{K}(\u_{xx}) +\alpha^2( [\mul{L}(\u),2\mul{L}(\u_x)+\mul{R}(\u_x)] 
+ [\mul{R}(\u),2\mul{R}(\u_x)+\mul{L}(\u_x)] )
+(\beta -4\alpha^2)\alpha \mul{K}(\u^3),
\end{gather}
where $\mul{K} = \mul{L}+ 2\mul{R}, -2\mul{L} -\mul{R}$, $\mul{L}-\mul{R}$,
using the left and right multiplication operators \eqref{L.R.mult}.

We can classify the types of evolution equations in this two-parameter family \eqref{mkdv.fam2}
by first considering the real reduction in which $\u$ and $\Psi$ get replaced by $u$ and $\psi$, 
which gives
$u_t = 3(\beta -4\alpha^2) u^2 u_x + u_{xxx}$. 
There are two cases, given by $\beta -4\alpha^2$ is zero or non-zero. 

In the case $\beta \neq 4\alpha^2$, 
the reduction yields the real KdV equation and its Lax pair with weight $w_L = 1$, 
up to a scaling transformation on $u$. 
Then, putting $3(\beta -4\alpha^2) = \sigma$ by scaling $\u$, 
we have
\begin{equation}
\beta = 4\alpha^2 + \tfrac{1}{3}\sigma,
\end{equation}
which gives 
\begin{equation}\label{mkdv.case3}
\u_t = (\tfrac{1}{3}\sigma -2\alpha^2 ) \{\u^2,\u_x\}
+(4\alpha^2 +\tfrac{1}{3}\sigma) \u\u_x\u 
+ \alpha [\u,\u_{xx}] + \u_{xxx} . 
\end{equation}
This is a one-parameter family of novel octonion evolution equations. 
It has as special cases: 
$\u_t = \u \u_x \u  + \tfrac{1}{\sqrt{6}} [\u,\u_{xx}] + \u_{xxx}$ 
for $\alpha = \tfrac{1}{\sqrt{6}}$ and $\sigma=1$; 
and 
$\u_t = -\tfrac{1}{2} \{\u^2,\u_x\}  + \tfrac{1}{2\sqrt{3}} [\u,\u_{xx}] + \u_{xxx}$ 
for $\alpha = \tfrac{1}{2\sqrt{3}}$ and $\sigma=-1$. 

In the case $\beta =4\alpha^2$, 
the real reduction is the linear Airy equation $u_t = u_{xxx}$. 
Taking 
\begin{equation}
\beta = 1, 
\quad
\alpha =\tfrac{1}{2}, 
\end{equation}
after scaling $\u$, 
we obtain 
\begin{equation}\label{mkdv.case4}
\u_t = [[\u,\u_x]\u] +\tfrac{1}{2} [\u,\u_{xx}] + \u_{xxx} . 
\end{equation}
This is another novel type of octonion equation, 
which relies on a lack of commutativity.

\subsection{Higher weight $w_\L$}

We repeat the search using higher weights $w_\L$. 
For $w_\L=2$, the results comprise two evolution equations: 
a previous octonion mKdV equation $\u_t = -\tfrac{1}{2} \{\u^2,\u_x\}  + \u_{xxx}$ 
with the Lax pair $(\L^2,\M)$ in terms of the $w_\L=1$ operators $\L$ and $\M$;
an octonion potential KdV equation
\begin{equation}
\u_t = \sigma\u_x^2 + \u_{xxx} , 
\end{equation}
which has $\L\Psi$ and $\M\Psi$ given by the operators
\begin{equation}
\L = \partial_x^2 + \tfrac{1}{3}\sigma \u_x, 
\quad
\M = 4\partial_x^3 + 2\sigma \u_x\partial_x + \sigma \u_{xx}
\end{equation}
acting by left and right multiplication.

\section{Concluding remarks}\label{sec:conclude}

All of the octonion integrable equations obtained here 
have quaternion counterparts 
simply by using quaternion multiplication in place octonion multiplication. 
This is a direct consequence of the Cayley-Dickson doubling structure 
relating the two algebras \cite{Conway-book}. 

A search for integrable octonion evolution equations 
having other scaling weights for $t,x,\u$ will be continued in future work. 
In particular, we will consider 5th order equations of 
KdV-type $t\to \mu^{-5} t$, $x\to \mu^{-1} x$, $u\to \mu^{2}u$, 
and mKdV-type $t\to \mu^{-5} t$, $x\to \mu^{-1} x$, $u\to \mu u$,
as well as Ibragimov-Shabat type $t\to \mu^{-6} t$, $x\to \mu^{-2} x$, $u\to \mu u$. 
An extension to include the complex-conjugate $\u$ variable will also be pursued. 

To facilitate the computations, 
the use of Maple will be replaced by a dedicated package in REDUCE 
for generating the ansatze by automatically excluding terms that vanish due to 
octonion product identities. 
This involves first finding all such identities with up to the maximum number of 
octonion variables appearing in the terms in the ansatze. 
Another package, CRACK \cite{Wol}, that runs in REDUCE will be used to solve the overdetermined system for the real coefficients in the ansatze.

\section{Acknowledgements}
S.C.A. and T.W. are each supported by an NSERC Discovery grant.

\appendix
\section*{Appendix: Octonion product identities}

Products of at most two octonions $\octa$ and $\octb$ are associative: 
\begin{gather}\label{prod.id1}\tag{A.1}
(\octa\octa)\octa = \octa(\octa\octa) = \octa^3, 
\quad
(\octa\octa)\octb = \octa(\octa\octb),
\quad
(\octa\octb)\octa = \octa(\octb\octa),
\quad
(\octb\octa)\octa = \octb(\octa\octa) . 
\end{gather}
These extend to all higher-degree products. 

Associator of three octonions is defined by $[\octa,\octb,\octc] := (\octa\octb)\octc - \octa(\octb\octc)$. 
This product is totally antisymmetric: 
\begin{equation}\label{prod.id3}\tag{A.2}
[\octa,\octb,\octc] = -[\octb,\octa,\octc] = -[\octa,\octc,\octb] . 
\end{equation}
No other identities exist involving products with degree three 
of at most three octonions.

\end{document}